\documentclass[twocolumn,8pt, showpacs, aps,  prd,  a4paper, superscriptaddress,  nofootinbib, tightenlines,  floats,floatfix]{revtex4}
\usepackage{graphicx}
\usepackage{epsfig}
\usepackage{caption}
\usepackage{latexsym}
\usepackage{amsmath,amssymb}
\usepackage{lipsum}

\newcommand{\beq}{\begin{equation}}
\newcommand{\eeq}{\end{equation}}
\newcommand{\beqa}{\begin{eqnarray}}
\newcommand{\eeqa}{\end{eqnarray}}
\newcommand{\bea}{\begin{array}}
\newcommand{\ena}{\end{array}}

\def\R{{\cal R}}

\def\be{\begin{equation}}
\def\ee{\end{equation}}
\def\bea{\begin{eqnarray}}
\def\eea{\end{eqnarray}}

\def\4pig{\sfrac{4\pi G}{c^{4}}}

\def\hsp5{\hspace{5mm}}
\newcommand{\sfrac}[2]{{\textstyle{#1\over#2}}}
\def\case#1/#2{\textstyle\frac{#1}{#2}}

\begin{document}
\title{ Holographic Entanglement Entropy  for noncommutative  Anti-de Sitter space }

\author{Davood Momeni}
\email{momeni_d@enu.kz,d.momeni@yahoo.com}
\affiliation{Eurasian International Center for Theoretical Physics and Department of
General \& Theoretical Physics, Eurasian National University, \\
Astana 010008, Kazakhstan}

\author{ Muhammad Raza }
\email{mreza06@gmail.com, mraza@zju.edu.cn} \affiliation{Department
of Mathematics, COMSATS Institute of Information Technology, Sahiwal
57000, Pakistan} \affiliation{State Key Lab of Modern Optical
Instrumentation,Centre for Optical and Electromagnetic Research,
Department of Optical Engineering, Zhejiang University, Hangzhou
310058, China}

\author{Ratbay Myrzakulov}
\email{rmyrzakulov@gmail.com}
\affiliation{Eurasian International Center for Theoretical Physics and Department of
General \& Theoretical Physics, Eurasian National University, \\
Astana 010008, Kazakhstan}

\date{\today}


\begin{abstract}
A metric is proposed to explore the noncommutative form of the Anti-de Sitter  space due to quantum effects. It has been proved that the
noncommutativity in AdS space induces a single component gravitoelectric field. The holographic Ryu-Takayanagi (RT) algorithm is
then applied to compute the entanglement entropy in dual $CFT_2$. This calculation can be exploited to compute UV-IR cutoff dependent central charge of the certain noncommutative $CFT_2$.  This non commutative computation of the entanglement entropy can be interpreted in the form of the surface/state correspondence. We have shown that non commutativity increases the dimension of the effective Hilbert space of the dual CFT.
\end{abstract}

\maketitle

\textit{Introduction}  The challenge now is to move the string
theory from "feasible" to "practical" in order to describe strongly
coupled quantum systems  in high energy and condensed matter
physics. The fundamental key of the path is a gauge-gravity pattern
called Anti-de Sitter space/Conformal Field Theory (AdS/CFT)
\cite{Maldacena}. Acknowledging the presence of AdS boundary of
gravitational bulk,  asymptotic values of certain fields could act as
dual quantum operators in CFT. This law will become a fundamental
part of that ever string/fields duality. We can actually perform
computations in CFT with the need for a suitable dictionary in AdS bulk.
From a common sense point of view, evidences suggest that we are
being modified significantly with quantum effects at Planck length,
$\lambda_p =  {({\hbar}G/c^3)  }^{1/2} \approx 1.6\times 10^{-33} cm$
  that modify geometry of spacetime. The Planck scale physics
offered to modify the Riemannian manifold to eliminate the
singularities after a quantum epoch. To provide efficient geometry
of the spacetime, noncommutative geometry (NCG) have been proposed
to deliver the quantum effects. There arise a fundamental question whether it is right in principle to modify Lorentz algebra(symmetry) generally?
 \cite{Snyder:1946qz}.  A. Connes formulated his now famous framework of $C^*$
algebras on spacetimes \cite{Connes:1994yd}.  This could include
deformation quantization of Poisson manifolds which were formulated
to replace lost parts \cite{Kontsevich:1997vb}.   These
noncommutativity of the coordinates of spacetime have been
successfully formulated to support and maintain  the anti-symmetric
tensor field arising from massless states of strings
 \cite{Seiberg:1999vs} :
\begin{equation}
[x^{\mu},x^{\nu}]
 = i \Theta^{\mu \nu} \label{commut}
\end{equation}
where $\Theta^{\mu \nu}$ denotes a constant skew-symmetric tensor.
The latter point has to be stressed since we did not address the
question of  Moyal product at all:
\begin{equation}
\displaystyle{(f\star
g)=f(x)exp\big(\frac{i}{2}}\Theta^{\mu
\nu}\overleftarrow{\partial_{\mu}}
\overrightarrow{\partial_{\nu}}\big)g(x) \,.
\end{equation}
While noncommutative spacetime with the commutation relation have
proven Lorentz violation over manifolds
\cite{Aschieri:2005yw,Aschieri:2005zs}, they are severely improved
via the twisted Poincar\'e algebra \cite{Aschieri:2005zs}. The most
suitable abelian twist element for this is the following:
\begin{equation}
{\cal{F}}= \exp\big(
{-i\over2} \Theta^{\mu \nu} \partial_{\mu}\otimes
\partial_{\nu} \big).
\end{equation}
The abelian twist elements are used in universal enveloping algebra
of the Poincar\'e algebra  describing the noncommutative
multiplication of our functions. A paper describing a fundamental
and systematic study of noncommutative Riemannian geometry (NCRG)
has been published  \cite{Chaichian:2006ht}. The  isometric
embeddings of a curved commutative spacetime in a flat higher
dimensional spacetime \cite{Friedman:1961, Nash:1956},  is a term
describing professional-algorithm that is extremely applicable. They
introduce some principles of good noncommutativity that must be
applied if embedding of commutative metric is adequately used.
NRG transformations may be applied to alter the shape of the known
metrics \cite{Wang:2009ju,Wang:2008ut,Romero-Ayala:2015fba}. We 've
extended the AdS/CFT even further and come up with the
 NCAdS/CFT \cite{Pramanik:2014mya}, gravity dual to NC
gauge theories \cite{Li:1999am} and even for Holography of NC
geometries (NCG)\cite{Chang:1999jma}.

One of most important  quantum measurement is entanglement entropy
(EE) of the dual quantum system with CFT via AdS/CFT principle. As
an example, consider the density matrix $\rho\equiv|\Psi><\Psi|$ for
the dual CFT system  in a pure quantum state $|\Psi>$. The EE of
systems never exist independent of the von Neumann entropy
 $S_{X}=-Tr_{X}(\rho_{X}\log\rho_{X})$. When $S_A$ is computed,
a complement observer $B$ might be inclined to say that remnant will
be calculated for the end value of the von Neumann entropy. This is
the sense in which the von Neumann entropy define "EE"
\cite{Calabrese:2005zw}. We were astonished to discover that EE has
become a geometrical, boundary entropy via AdS/CFT \cite{hee1,hee2},
called holographic entanglement entropy (HEE) (see for a review
\cite{Nishioka:2009un}).  In this scheme, you'll discover the surface
area entropy, one of magic's most enduring classics
 in gravitational physics \cite{Srednicki:1993im}. This has been used
recently to produce a quantitative model for critical phenomena with
some success \cite{Peng:2014ira}-\cite{Romero-Bermudez:2015bma}. \par
 Quantum entanglement can be used as an alternative candidate
for dark energy, an exotic fluid with negative equation of state
(EoS), which is believed to be responsible for current acceleration
of the Universe
\cite{Capozziello:2013sfa}-\cite{Capozziello:2010yq}. Particularly,
it is possible to describe the phase transition from early epoch to
the late time as two entangled cosmological eras. In this approach,
the aim is to describe acceleration of the Universe, by measuring
the entanglement degrees of freedom \cite{Capozziello:2014nda}.
Furthermore, there is a consistent way to address the cosmic
acceleration using the entanglement measures by using an EoS which
is calculated  from the existed quantum entanglement  between
different cosmological epochs \cite{Capozziello:2013wea}. This is a
novel approach to address dark energy as biproduct state of the
quantum entangled subsystems (here early and late time epoches).
\par
A
major goal of this letter  is to understand how HEE deformed on
$NCAdS_3$ background.
Non-commutative computation of the HEE has been investigated by
several authors. For example in Ref. \cite{Fischler:2013gsa}, the
authors investigated HEE  in a large-N strongly coupled
noncommutative gauge theory at zero and finite temperature regimes.
Using the RT duality, they assumed that this NC gauge theory in
boundary has a "fully commutative" AdS bulk geometry. So, in one
side they have NC gauge theory as an entangled quantum system and in
other side they have a commutative AdS bulk geometry. The extremal
surface is also assumed to be commutative. Non-commutativity exists
only at the level of quantum boundary system. They found a
monotonic-increasing form of HEE with respect to the length, in the
system.  Furthermore, in Ref. \cite{Karczmarek:2013xxa}, a
holographic computation of HEE was done in  strongly coupled non-
local dual  field theories as a noncommutative deformation of SYM
theory. They supposed that the gravitational dual is a pure-AdS,
because all the cases which they studied were about UV deformations
of the $N = 4$ SYM. Also in this work the authors attacked to the
non-commutativity from the boundary quantum field theory point of
view. Their geometry  still  remains commutative. \par But our
problem in this letter is different from the above references. When
there is a fully satisfactory version of NC bulk geometry, we have
two major options to treat HEE from the AdS/CFT approach. One is to assume
that this NC geometry has a "commutative" dual CFT. Other choice is
to suppose that the dual CFT on boundary is commutative.
What we
proceed in this letter was to suppose that \emph{NC-AdS geometry in
bulk has a NC dual CFT on boundary}. What we computed is the central
charge (Casimir energy) of the $NC-CFT_2$ using $NCAdS_3$ bulk
geometry via RT algorithm.
However we don't know the structure and
the correct form of such hypothesis $NCCFT_2$, but we are lucky.
  Our
main  motivation is to
compute the NC-central charge of this quantum theory using the associated HEE  for $NCAdS_3$.
These
holographic computations have been interpreted as an appeal to the
leakage of the NC-AdS/CFT, but
This assumption is entirely out of
keeping with all we know about NC gauge theories
\cite{Karczmarek:2013xxa}, over whom NC says it practically ruled as
a NC-SYM gauge theory. This letter gives examples of NC geometries
interpreted by the NC-CFT.

 We proposed to explore the noncommutative
form of the AdS space via \cite{Chaichian:2006ht}. The holographic
$AdS_3/CFT_2$ algorithm \cite{hee1,hee2} is then used to compute the
HEE in dual $CFT_2$ in commutative boundaries.
\par
\textit{Foundations of NCRG} Due to standard RG convention each
element of the $(N^{1,n-1}, g)$ = $n$-dimensional Lorentzian
manifold  is labeled in the most comprehensible manner for the
metric $g$ with signature $(-1,1,\cdots,1)$\cite{Chaichian:2006ht}.
NCRG  is achieved by enriching and extending the Nash's isometric
embedding of pseudo-Riemannian manifolds. Extending this just a
little further allows definitions with a set of smooth function $X^1,
\cdots, X^p, X^{p+1},\cdots, X^{p+q}$ on $N^{1,n-1}$, such that
 \begin{eqnarray}
g=-\Sigma_{a=1}^p\big(dX^a\big)^2
+\Sigma_{b=p+1}^{p+q}\big(dX^{b}\big)^2\label{Nashg}.
\end{eqnarray}
It helped define the coordinate chart $U$ of $N^{1,n-1}$ and natural
coordinates =$\{x^0, x^1, \cdots, x^n\}$ for it. How would we define
NC behavior in metric?.  Objectives need to define a specific
parameter $\bar{h}$ which is a definite and real indeterminate
parameter. In this letter, we define ring of formal power series in
$\bar{h}$ in the more conventional manner, using the shorter list
$\R[[\bar{h}]]$. We can always define $\bar{h}$ in our formalism as
the ratio of the standard model mass scale and the Planck mass of
course. Suppose a set  $A$ of formal power series in $\bar{h}$ with
coefficients always $\in\mathcal{R}$. We are looking to expand any
arbitrary element of $A$ with a summation form of $\sum_{i\ge 0}
f_i\bar{h}^i$ where  $f_i$ are smooth functions on $U$.  It is clear
to us from a reading of these steps that  $A$ means
$\R[[\bar{h}]]$-module. One can thus postulate that Moyal product $u\star v$ is a map on $U$  far as
it maintains a highly centralized role of NCRG:
\begin{eqnarray}\label{multiplication}
(u\star v)(x) = \lim_{x'\rightarrow x}\ \exp{\left(\bar{h} \sum_{i
j} \theta_{i j}\partial_i\partial_j^\prime\right)}u(x)
v(x')\label{star}.
\end{eqnarray}
We notate $\partial_i=\frac{\partial}{\partial x^i}$,  and $(\theta_{i
j})$ is a constant skew symmetric $n\times n$ matrix. Moyal product
has managed to preserve the associativity and Leibnitz rule of its
formal form:
$$\partial_i(u\ast v)= \partial_i u\ast v + u\ast \partial_i v,\ \ m=p+q\in\mathcal{Z}^{+}$$
It is wise to obtain a copy of dot-product for an element
\begin{eqnarray}\label{dotproduct-dimm}
A=(a_1, \dots, a_m): A^m\otimes_{\R[[\bar{h}]]} A^m\longrightarrow
A^m
\end{eqnarray}
on $U$ is defined by:
\begin{eqnarray*}
A\bullet B=-\sum _{i=1} ^p a_i\star b_i +\sum _{j=p+1} ^{p+q} a_i
\star b_i
\end{eqnarray*}
We can always define  the NC metric in terms of  functions
 $X\in A^m$,
 $E_i=\partial_i X$:
\begin{eqnarray}
\hat{g} _{i j}=E_i \bullet E_j\label{ncg}.
\end{eqnarray}
We denote these metrics as " NC" non-singular $\hat{g}=(\hat{g}_{i j})$ the $n\times n$ matrix with entries $\hat{g}_{i
j}\in A$.
Let  $(\hat{g}^{i j})$ denote the standard  inverse :
\begin{eqnarray*}
\hat{g} _{ij} \star \hat{g} ^{jk} =\hat{g} ^{k j}  \star \hat{g} _{j i} =\delta _i^k \label{ghat}.
\end{eqnarray*}
Results will define the connection of the metric in response to spacetime deformation:
$$\nabla_{i}E_{j}= \Gamma_{ij}^{k}\star E_{k},\label{coneccionNC}$$
here $\Gamma_{ij}^{k}= \Gamma_{ijl}\star{g}^{lk}$ and
\begin{equation}
\Gamma_{ijk}= \partial_{i}E_{j}\bullet E_{k}.
\end{equation}
and the associated noncommutative Riemann and Ricci scalars are
defined as:
\begin{eqnarray}
&&R_{kij}^{l}=
\partial_{i}\Gamma_{jk}^{l}-\partial_{j}
\Gamma_{ik}^{l}+\Gamma_{jk}^{p}
\star\Gamma_{ip}^{l}-
\Gamma_{ik}^{p}\star\Gamma_{jp}^{l},\\&&
R_{j}^{i}={g}^{ik}\star R_{kpj}^{p},\ \
\Theta^{l}_{p}={g}^{ik}\star R^{l}_{kpi},\ \  R=R_{i}^{i} .
\end{eqnarray}
In our case the appropriate addressee of Einstein equations with cosmological constant from the auditor to those modifications with NCRG is the following:
\begin{equation}
R_{j}^{i}+
\Theta_{j}^{i}-\delta_{j}^{i}R+2\delta_{j}^{i}\Lambda=2T_{j}^{i}, \label{NCEinsteinEq}
\end{equation}
 where $T_{j}^{i}$ denotes the generalized \emph{energy-momentum} tensor, $\Lambda$
is the cosmological constant. Exact results have been reported in a
few papers \cite{Wang:2009ju,Wang:2008ut,Romero-Ayala:2015fba}. The
goal of next section is to realize NC$AdS_3$ via the mentioned
formalism.
\par
\textit{Embedding technique }:
To find out more about the form on NCAdS follow the embedding scheme for
 $AdS_3$ spacetime. This showed that  $AdS_3$ were already embedded into the $4$D flat spacetime $\mathcal{R}^{2,2}$:
\begin{eqnarray}
g=-\big(dX_1^2+dX_2^2\big) +\big(dX_3^2+dX_4^2\big).\label{hyper}
\end{eqnarray}
We will be comparing (\ref{hyper}) with (\ref{Nashg}). We observe
that $p=q=2$. This embedding  is  invariant  under $SO(2,1)$,
which is the precise isometry group of AdS, with conformal boundary.
An appropriate rescaling casts the boundary in hyperboloid form  in
$(2)D$ form, which is universal for systems supporting $SO(1,2)$
isometry group, with one dilatation and two special conformal
transformations. Now, we find the embedding coordinates
$\big(t,\rho,\theta\big)$ for all of the hyperboloids:
\begin{eqnarray}
&&X_1=l\cosh\rho\sin t,\\&&
X_2=l\cosh\rho\cos t,\\&&
X_3=l\sinh\rho\sin\theta,\\&&
X_4=l\sinh\rho\cos\theta
\end{eqnarray}
We need to specify an AdS radius, for this we have to define it by
$l^2=-\frac{3}{\Lambda}$. These coordinates represent a special
universal covering  of the commutative $AdS_3$ spacetime:
\begin{eqnarray}
g=l^2\big(-\cosh^2\rho dt^2+d\rho^2+\sinh^2\rho
d\theta^2\big)\label{unig}.
\end{eqnarray}
The domain $\rho=0$, represent the the AdS boundary of the AdS
metric. The aim of the next section is to compute the noncommutative
version  of  (\ref{unig}) from (\ref{ncg}).
\par
\textit{Explicit form of NC$AdS_{3}$}: Evaluation techniques of
(\ref{ncg}) include star product and skew symmetric matrix
$\theta_{ij}$. The starting point for the computation of (\ref{ncg})
is the functions  multiplying (\ref{star}), which we can write in
the following simple form:
\begin{eqnarray}
&&A(y)\star B(y')=\lim_{y\to y'}e^{\bar{h}\big(\partial_\rho\partial
_{\theta'}
-\partial_{\theta}\partial_{\rho'}\big)}A(y)B(y'),\label{star}\\&&\nonumber
y\equiv(t,\rho,\theta),
\end{eqnarray}
where
$$
\big(\theta_{ij}\big)_{3\times 3}=\left[ \begin {array}{ccc} 0&0&0\\ \noalign{\medskip}0&0&1
\\ \noalign{\medskip}0&-1&0\end {array} \right].
$$
The computation of the (\ref{ncg}) is performed continuously for all
orders of $\bar{h}$, using the Zassenhaus  formula\cite{Zassenhaus}:
\begin{eqnarray}
&&e^{\bar{h}\big(\partial_\rho\partial
_{\theta'}
-\partial_{\theta}\partial_{\rho'}\big)}=e^{\bar{h}\big(\partial_\rho\partial
_{\theta'}\big)}e^{-\bar{h}\big(
\partial_{\theta}\partial_{\rho'}\big)}
\end{eqnarray}
 This allows routine computation of the components  of NCAdS.
Here is the complete form of the nonvanishing components of the noncommutative version of $AdS_3$ spacetime:
\begin{eqnarray}
&&\hat{g}_{tt}=-l^2\cosh^2\rho\label{g11}
\\&&
\hat{g}_{\rho\rho}=l^2+2l^2\sin^2(\frac{\bar{h}}{2})\Big(1-\cos\bar{h}\cosh(2\rho)\Big)\label{g22}
\\&&
\hat{g}_{\rho\theta}=l^2\sin^2(\frac{\bar{h}}{2})\Big(e^{2\rho}\sin\bar{h}
-\sin(2\theta-\bar{h})\\&&\nonumber+e^{-2\rho}\sin(2\theta)\Big)
\label{g23}
\\&&
\hat{g}_{\theta\theta}=l^2\sinh^2\rho+\frac{l^2}{\sqrt{2}}\sin^2(\frac{\bar{h}}{2})\Big[e^{2\rho}\cos\bar{h}-\frac{1}{\sqrt{2}}\\&&\nonumber-e^{-2\rho}\cos(\bar{h}+\frac{\pi}{4})+2\sinh(\rho)\cos(2\theta-\frac{\pi}{4})
\\&&\nonumber+2\cos(2\theta-\frac{\pi}{4})\cos\bar{h}\Big]\label{g33}
\end{eqnarray}
The metric components are reduced to a standard $AdS_3$ in limit
with $\bar{h}=0$. Deformation of (\ref{unig}) using
(\ref{g11},\ref{g22},\ref{g23},\ref{g33}) is given by the following
metric:
\begin{eqnarray}
\hat{g}=g+\big(\delta  g_{\rho\rho}d\rho^2+2 \delta
g_{\rho\theta}d\rho d\theta+ \delta
g_{\theta\theta}d\theta^2\big)\label{ncmetric}.
\end{eqnarray}
where $\delta g_{ij}\equiv |\hat{g}_{ij}-g_{ij}|\ll g_{ij}$. There
is a \emph{dimensional reduction} in noncommutative part of metric.
It was the first observation in the NCRG to use noncommutativity to reduce
dimensions of the commutative bulk and is fully surprising. If
Planck scale quantum effects are sufficient to deform the
commutative bulk, then noncommutativity is able to reduce bulk's dimensions will usually
be arisen. Ascending to the NC  metric  (\ref{ncmetric}), the
portion of noncommutativity escaping from the purely $\rho$
dependency  to the $(\rho,\theta)$ dependent metric in one lower
dimension. The NC portion of the metric gets sent to the spacelike
metric due to the absence of $\delta{g}_{tt}$. The  deformed metric
can write even more wellness to produce a gravitoelectric field
\cite{landau}:
\begin{eqnarray}
E_g\equiv -\frac{1}{2}\nabla_{\gamma_{ab}}\log {g}_{11}.
\end{eqnarray}
where we define the length of a strip in the Euclidean coordinates
$\bar{\rho}=i\rho,\phi=-i\theta$:
\begin{eqnarray}
&&\gamma\equiv (-{g}_{11}\tilde{A}^2+g_{22})d\phi^2.
\end{eqnarray}
where
\begin{eqnarray}
&&g_{11}=\delta g_{\rho\rho}|\{\rho\to i\tilde{\rho},\theta\to-i\phi\},\\&&g_{22}=\frac{\delta g_{\theta\theta}}{\delta g_{\rho\rho}}\{\rho\to i\tilde{\rho},\theta\to-i\phi\},\\&&
\tilde{A}=\frac{\delta g_{\rho\theta}}{\delta g_{\rho\rho}}\{ \rho\to i\tilde{\rho},\theta\to-i\phi\}
\end{eqnarray}
There is no doubt about a fact that {\it noncommutativity in AdS
spsacetime induces a single component gravitoelectric field } .

\par
\textit{HEE for $NCAdS_3$:}
The HEE is defined as the entropy of a region of space $\tilde{A}$ and its complement  on the minimal surfaces in $AdS_{d+1}$ \cite{hee1,hee2}:

\begin{eqnarray}
S_{\tilde{A}}\equiv S_{HEE}=\frac{Area(\gamma _{\tilde{A}})}{4G_{d+1}}.\label{HEE}
\end{eqnarray}%
What we need is to compute the  $(d-1)$D minimal surface $\gamma _{%
\tilde{A}}$ on equal time patches. We suppose that we can extend
$\gamma _{\tilde{A}}|_{AdS_{d+1}} $ inside the bulk, and we are also
limited, to a certain boundary condition, by keeping the boundaries
same  $\partial \gamma _{\tilde{A}}=\partial {\tilde{A}}$. Several
options for the parametrization of the boundary surfaces are
available as well as different choices for the form of minimal
surface functional in (\ref{HEE}). Here the assumption for HEE
purpose is that none of the minimal surfaces on the boundaries  is
noncommutative under (\ref{commut}). We consider that the assumption
of commutativity is correct and may validate some of the arguments.
Often the implicit assumption was that minimal surface appeared into
a strip from which the UV boundary of AdS ever cutoffed by
$\rho=\rho_0$:
\begin{eqnarray}
&&
\tilde{A}%
:=\{t=t_{0},0\leq \theta \leq \theta_0),\rho=\rho(\theta
)\},\theta_0=\frac{2\pi l}{L},
\end{eqnarray}
here $L$ denotes the spacial length of the total system which  indeed, is infinitely long and $l$ denotes the length of a system $\tilde{A}$, formally $L\gg l$.
The HEE expression (\ref{HEE}) can then be corrected by inverting the affected  NCAdS metric given by  (\ref{ncmetric}):
\begin{eqnarray}
&&S_{HEE}^{NCAdS_3}=S_{HEE}^{AdS_3}-\frac{1}{2\sqrt{\pi}}\Sigma^{\infty}_{nmk},a_{nmk}
\Delta S_{HEE}^{nmk},\label{SNC}\\&&
a_{nmk}=\frac{(n-3/2)!}{n!m!(n-k)!(k-m)!}, 1\leq m \leq k \leq n.
\end{eqnarray}
here $S_{HEE}^{AdS_3}=\frac{l}{4G_3}\log\Big(e^{2\rho_0}\sin\big(\frac{\pi l}{L}\big)\Big)$ was calculated in \cite{hee1} in terms of central charge of $(1+1)$ CFT and geometrical lengths of spacelike sector of AdS spacetime. Here $a$ stands out for an ultraviolet (UV) cutoff.
Consequently, what major has been done is to write an expression for NCHEE as follows:
\begin{eqnarray}
&&\Delta S_{HEE}^{nmk}=\frac{1}{4G_3}\int_{0}^{\theta_0} d\theta\Big[ (\rho')^{2m+n-k}\\&&\nonumber\big(l^2\rho'^2+l^2\sinh^2\rho\big)^{\frac{1}{2}-n}
\big(\delta g_{\rho\rho}\big)^m\big(2\delta g_{\rho\theta}\big)^{n-k}\big(\delta g_{\theta\theta}\big)^{k-m}\Big].
\end{eqnarray}
We define a Lagrangian density as the following:
\begin{eqnarray}
&&\mathcal{L}^{nmk}(\rho,\theta)\equiv\Big[ (\rho')^{2m+n-k}\big(l^2\rho'^2+l^2\sinh^2\rho\big)^{\frac{1}{2}-n}\\&&\nonumber\times
\big(\delta g_{\rho\rho}\big)^m\big(2\delta g_{\rho\theta}\big)^{n-k}\big(\delta g_{\theta\theta}\big)^{k-m}\Big].
\end{eqnarray}
Unfortunately, this Lagrangian does not satisfy Beltrami's
identity.\footnote{It stated: these Lagrangians do not satisfy a
simple identity in the form
$\mathcal{L}-\rho'\partial_{\rho'}\mathcal{L}\neq C$. The only
Beltrami's case came from $m=n=k$ term, where the geodesic
$\rho(\theta)$ obtained from (\ref{geodesic})  was not analytical. }
The static geodesic satisfies the following equation:
\begin{eqnarray}
\partial_{\theta}\big(\partial_{\rho'}\mathcal{L}^{nmk}\big)
-\partial_{\rho}\mathcal{L}^{nmk}=0.\label{geodesic}
\end{eqnarray}
Our problem is to minimize the following functional:
\begin{eqnarray}
&&Minimize\{I^{nmk}[\rho(\theta)]=\int_{0}^{\theta_0} d\theta
\mathcal{L}^{nmk}(\rho,\theta)\},
\end{eqnarray}
that connects the boundary points $\rho(0)=\rho(\theta_0)=\rho_0\gg
1$. Leading the $m=n=k=1$ of integral to look at $\Delta
S_{HEE}^{nmk}$ is  really appropriate here. An ideal solution here
is to choose a phase-space like solution for (\ref{geodesic}) in the
form $\rho=\rho'(\rho)$ :
\begin{eqnarray}\label{rho'}
&&\rho'^2=\frac{E^2 f^2\sinh^4\rho }{3}-\sinh\rho\\&&\nonumber+
\left\{
\begin{array}{lr}
-2\frac{|q|}{q}\sqrt{-\frac{p}{3}}\cosh\big(\frac{1}{3}\cosh^{-1}\big(-\frac{3|q|}{2p}\sqrt{-\frac{3}{p}}\big)\big)\ , & p<0 \\
-2\sqrt{\frac{p}{3}}\sinh\big(\frac{1}{3}\sinh^{-1}\big(\frac{3q}{2p}\sqrt{\frac{3}{p}}\big)\big)\ , & p>0%
\end{array}%
\right. \ .
\end{eqnarray}%
Suppose $f\equiv
2l^2\sin^2(\frac{\bar{h}}{2})\Big(1-\cos\bar{h}\cosh(2\rho)\Big)$
and $4p^3+27q^2>0$, we define :
\begin{eqnarray}
&&p=-\frac{E^2 f^2\sinh^5\rho }{3}\big(E^2f^2\sinh^3\rho-6\big),\\&&
q=-\frac{E^2f^2\sinh^6\rho}{27}\big(2E^4f^4\sinh^6\rho\\&&\nonumber-18E^2f^2\sinh^3\rho
+27\big)
\end{eqnarray}
We observe that $p<0,q<0$ in our model.

However rewriting $\Delta S_{HEE}^{nmk}$ in terms of $\rho$  really
produce exceptionally pretty links between $\rho,\rho'$:

\begin{eqnarray}\label{111}
&&\Delta S_{HEE}^{111}=\frac{l}{2G_3}\sin^2(\frac{\bar{h}}{2})\lim_{\epsilon\to0}\int_{\epsilon}^{\rho_0}d\rho \Big[\rho'
\\&&\nonumber\times\big(\rho'^2+\sinh^2\rho\big)^{-\frac{1}{2}}\big(1-\cos\bar{h}\
cosh(2\rho)\big)\Big] .
\end{eqnarray}
The phase-space solution (\ref{rho'}) is used to evaluate the
difference (\ref{111}) between pure commutative AdS  and NC one. The
numeric integration will be applied to evaluate  (\ref{111}) the
effect of various forms of NC terms:
\begin{eqnarray}
&&\Delta S_{HEE}^{111}\simeq-\frac{l\gamma_1}{8G_3}\bar{h}^{8/3}\lim_{\epsilon\to0}\int_{\epsilon}^{\rho_0}d
\rho(\sinh\rho)^{3/4}\\&&\nonumber\times\big(
\cosh(2\rho)-1\big)^{11/6}.
\end{eqnarray}
here
$\gamma_1=16\sqrt{6}l^{10}E^5,E=\mathcal{L}^{111}-\rho'\partial_{\rho'}\mathcal{L}^{111}
\equiv\emph{constant}$. Finally we compute the "physical" (the $\Re$
part) of  the leading integral, we obtain:
\begin{eqnarray}
&&\Delta S_{HEE}^{111}\simeq -\frac{3.95l}{G_3}
\Big[(l^{2}E)^5\bar{h}^{8/3}\Big]{\rm e}^{\rho_{0}}. \label{final}
\end{eqnarray}
Finally we obtain the NC corrected form of HEE for $AdS_3$:
\begin{eqnarray}
&&S_{HEE}^{NCAdS_3}=\frac{c_{NC}}{3}\log\Big(e^{2\rho_0}\sin\big(\frac{\pi l}{L}\big)\Big),\\&&
c_{NC}=c+\frac{23.7(l^2E)^5\bar{h}^{8/3}}{\log\big(\frac{L}{a}\sin\big(\frac{\pi l}{L}\big)\big)}\frac{L}{a}
\end{eqnarray}

We found a mathematical formula that worked much more than anything
capable of becoming the basis for the HEE of a $NCAdS_3$. The first
term in this formula becomes divergent completely in agreement to
the commutative case of $AdS_3$, it is safe to use on all higher
dimensional extensions. The HEE for a $NCAdS_{d\geq4}$ can be
calculated in an analogous manner to the $AdS_3$ in NCRG. The first
term, known as \emph{noncommutative area} law, presents the ruling
in holographic formula beginning if the spacetime becomes
noncommutative just slightly. However, it  manages to be surprising
or deviate from the  Ryu-Takayanagi formula of the \cite{hee1}. In
our NC case, the growth in the value of HEE is proportional to the
instantaneous value of the HEE. The rate may be positive or
negative. By comparing expressions in Ryu-Takayanagi formula and NC
case, we discover some possible perspectives in interpretation of
NCAdS spaces. \par
 If we substitute (\ref{final}) in (\ref{SNC})
we obtain:
\begin{eqnarray}
S_{Corrected}=\frac{\rho_0}{2G_3}+\frac{\mathcal{N}}{G_3}e^{\rho_0}+\text{constants}
\end{eqnarray}
here $\mathcal{N}=\frac{3.95l}{G_3}
\Big[(l^{2}E)^5\bar{h}^{8/3}\Big]$. Following a very important new
proposal for HEE as entropy of a quantum system in {\emph
surface/state} duality \cite{Miyaji:2015yva}, the above expression for $S_{Corrected}$ can be interpreted
as the  "effective" entropy  $S_{eff}$ or  the generalized holographic
entropy of a surface in bulk $\Sigma$. This entropy must be equal to
the $ \log[dimH_{eff}] $, which the $dimH_{eff}$ is introduced as
the effective dimension of the Hilbert space for the dual CFT \footnote{For an alternative definition of quantum field theory in non commutative systems see \cite{Chaichian:1998kp}.}
Because the non commutativity increased the amount of the entropy
enclosed in the bulk region $\Sigma$ (much more bigger than the
commutative spacetime due to the presence of the $\log$ term), it
means that the effective dimension is increased by non
commutativity. An emergence of the new degree of freedoms is
appeared. The order of change in the $dimH_{eff}$ for non
commutative system in the comparison to the commutative system is
about:
\begin{eqnarray}
\frac{dimH^{NCAdS_3}_{eff}}{dimH^{AdS_3}_{eff}}\simeq e^{\frac{\mathcal{N}}{G_3}e^{\rho_0}}\gg1
\end{eqnarray}
 which is a very large number of degrees of freedom. We
mention here that this result is based on a first order
approximation. So, the $dimH_{eff}$ may will change if we take into
account other higher order terms.

By comparing results from the both studies, we hope to
understand more about the impact of  HEE being reared in Planck
scale.

\par

\textit{Summary}
This letter offers a glimpse of how one can use holography to realize entanglement entropy for the noncommutative AdS space.
 We first find the  noncommutative Anti-de Sitter metric using a fully consistent version of
noncommutative Riemannian geometry. Later we use the Ryu-Takayanagi
formula to compute holographic entanglement entropy of a \emph{commutative}
region, needs to derive an expression for entropy of a possible
noncommutative CFT. A correction for the entanglement entropy  was
applied to the minimal case $n=m=k=1$ value because this was
generally large compared to the other higher terms. Planck scale
correction for entropy of noncommutative system of order
$\bar{h}^{8/3}$ contained the following parameters: a constant of
motion $E$ appeared as $E^5$, the powers of AdS radii $l$ as
$l^{11}$ and the Newtonian constant $G_{3}^{-1}$.  Collectively these calculations can compute  the central charge  for noncommutative $CFT_2$.
\par
  We might mislead ourselves if we interpreted this expression
as referring to cutoff independent central charge; on the other
hand, we more than most of the NC-SYM thinks of $c_{NC}$ as an
imitation of $NCAdS_3$. This central charge has been constituted for
a peculiar group of NC-AdS bulks, which have been interpreted as the
gravitational duals of the skeletons of $CFT$ of an NC boundary
conformal field theory. All these points of structure can only be
correctly interpreted after a consideration of the needs of the
individual NC-$CFT_2$, and of the large colony of which they are
members. We have seen that the $c_{NC}$ is preceded by the formation
of NC bulk geometry, and its appearance is now interpreted as a sign
of $CFT_2$ manufacture. This holographic computation has been
interpreted as an ad-hoc first stepping for a NC-AdS/CFT programm.
Whatever their origin is, these calculations tend to be interpreted as
"gift" to the AdS/CFT.
 Recommendation  is
to allow noncommutative corrections of $\Delta S_{HEE}^{nmk}$ that
requires further research.

\end{document}